\documentclass[12pt]{article}
\usepackage{amsmath}
\usepackage{psfig}
\begin{document}

\title{Assessing Turbulence Strength via Lyaponuv Exponents}
                   
\author{Mayer Humi\\
Department of Mathematical Sciences\\
Worcester Polytechnic Institute\\
100 Institute Road\\
Worcester, MA  01609}

\maketitle
\thispagestyle{empty}

\begin{abstract}
In this paper we study the link between `turbulence strength' in a flow
and the leading Lyaponuv exponent that characterize it. To this end we
use two approaches. The first, analytical, considers the truncated
convection equations in 2-dimensions with three (Lorenz model) and six
components and study their leading Lyaponuv exponent as a function of the
Rayleigh number. For the second approach we analyze fifteen time series 
of measurements taken by a plane flying at constant height in the upper 
troposphere. For each of these time series we estimate the leading Lyaponuv 
exponent which we then correlate with the structure constant for the 
temperature.
\end{abstract}

\newpage

{\bf Turbulence is still considered as one of the major unsolved problems of
modern science. Usually one attempts to characterize the turbulent state 
of the system in terms of various numbers such as the Reynolds number,
Grashof number, Rayleigh number, etc. How a combination of these numbers 
actually characterize the strength of turbulence in the system is
not answered easily. The situation is even more complex when the flow is
represented only by a time series. Furthermore is there any relation 
between this `strength' and chaos theory?  Although chaos owes it modern 
origins to the work of Lorenz who investigated the onset of convection, 
a characterization of turbulence in terms of dynamical invariants such
as Lyapunov exponents, fractal dimension and so on is still an open
question which needs further investigation. This paper attempts to 
present a modest contribution in this direction.}

\newpage

\section{Introduction}

Atmospheric convection in two dimensions is described [1,2] by the following 
system of partial differential equation:
\begin{equation}
\label{1}
\frac{\partial}{\partial t} \nabla^2\psi = -
\frac{\partial(\psi,\nabla^2\psi)}{\partial(x,z)} + \nu \nabla^4\psi +
\alpha g \frac{\partial\theta}{\partial x}
\end{equation}
\begin{equation}
\label{2}
\frac{\partial\theta}{\partial t} =
-\frac{\partial(\psi,\theta)}{\partial(x,z)} + \frac{\Delta T}{H}
 \;\frac{\partial\psi}{\partial x} + \kappa\nabla^2\theta
\end{equation}

Subject to the stress free boundary conditions:

\begin{equation}
\psi=0,\,\,\nabla^2\psi=0,\,\,\theta=0 ,\,\,\,\, z=0,1
\end{equation}
In this system $\psi$ is the stream function and $\theta$ is the potential
temperature. The constants $\alpha, \nu, \kappa, g$ denote respectively
the coefficient of thermal expansion,the kinematic viscosity,the thermal
conductivity and the acceleration of gravity. $H$ is the fluid layer 
thickness
and $\Delta T$ is the temperature difference between the upper and lower
surface of the fluid (which is assumed to be held constant). We also have
\begin{equation}
\label{3}
\frac{\partial(f,g)}{\partial(x,z)} = \left| \begin{array}{cc}
\frac{\partial f}{\partial x}    &\frac{\partial f}{\partial z}\\
\frac{\partial g}{\partial x}   &\frac{\partial g}{\partial z}
\end{array}
\right|
\end{equation}

In his famous 1963 paper [2] Lorenz introduced and studied a model in which
the solution to eq.(\ref{1})-(\ref{2}) is approximated by a three Fourier modes
\begin{equation}
\label{4}
\psi = \frac{\kappa(1+a^2)\sqrt{2}}{a}X(t)\sin\left(\frac{\pi
ax)}{H}\right) \sin\left(\frac{\pi z}{H}\right)
\end{equation}
\begin{equation}
\label{5}
\theta = \frac{R_c}{\pi R_a}\left\{\sqrt{2} Y(t) \cos\left(\frac{\pi a
x}{H}\right)\sin\left(\frac{\pi z}{H}\right) -
Z(t)\sin\left(\frac{2\pi z}{H}\right)\right\}
\end{equation}
where $a$ is a parameter and
\begin{equation}
R_a = \frac{g \alpha H^3 \Delta T}{\kappa \nu}
\end{equation}
\begin{equation}
R_c = \frac{\pi^4 (1+a^2)^3}{a^2}
\end{equation}
are the Rayleigh number and the critical Rayleigh numbers for the flow.
This led to the following three coupled equations for $X,Y,Z$.
\begin{eqnarray}
\label{6}
\dot{X} &=& -\sigma X + \sigma Y\\ \nonumber
\dot{Y} &=& XZ + rX - Y\\ \nonumber
\dot{Z} &=& XY - bZ
\end{eqnarray}
where
\begin{equation}
\label{7}
\sigma = \frac{\nu}{\kappa},\,\, b = \frac{4}{(1 + a^2)}, \,\, 
r = \frac{R_a}{R_c}.
\end{equation}

These equations are usually referred to as the `Lorenz model'. Since its
appearance 
this model and its implications were studied in great detail in hundreds of 
publications with special attention to its bifurcations as a function 
of the parameters $\sigma, r$ and $b$. It has been recognized [3] 
however that (as expected) the approximation of the solution to eqs 
(\ref{1}) and (\ref{2}) which is provided by (\ref{3})-(\ref{7}) becomes 
`poor' as $r$ increases.  This has 
led several authors to develop and study models with a larger number of 
modes [4,21]. 

The basic conjecture that we want to validate in this paper, through some
case study, is that stronger turbulence [22] in a system is linked to larger 
value of the leading Lyaponuv exponents [5,6] for the the flow. 
To test this conjecture we
study the Lyaponuv exponents of two truncated systems that were derived form 
eqs (\ref{1})-(\ref{2}). We then use Rosenstein algorithm [7] to
estimate the 
Lyaponuv exponents for fifteen atmospheric time series that were obtained
from measurements in the upper troposphere (10Km approximately).
The Lyaponuv exponent is then correlated with the structure constant 
for the temperature [8,9] which is representative of the 
density fluctuations
in the atmosphere and hence with the strength of turbulence in the flow.

To begin with we study the dependence of the leading
Lyaponuv exponent on the parameter $r$ in the Lorenz model. This parameter
is representative of the Rayleigh number in eqs. (1.1)-(1.2).
We show that for small values of $r$ the leading Lyaponuv exponent in 
the Lorenz model is either negative or small. Increasing
$r$ leads at the `onset of chaos' to a steep increase in the value of
this exponent. Further increase in $r$ leads to further modest increase 
in the value of the leading Lyaponuv exponent (which is indicative of the
fact that these modes are 'saturated').

One expects to obtain a better approximation to the solution of eqs.
(\ref{1})-(\ref{2}) as the number of modes allowed in the model increases. 
Hence as a second step in our study we consider a model with six modes.
The leading Lyaponuv exponent for this model displays a similar behavior
as for the Lorenz model. However the highest value of the leading Lyaponuv 
exponent in this model is larger than the one attained by the Lorenz model. 
This is consistent with our conjecture since a model with a larger number 
of modes can represent stronger turbulent state of the system due 
to the larger number of interacting modes. 

From a practical point of view it is important to assess turbulence
strength in a flow from a representative time series of measurements.
In this case although the equations governing the flow are well known
one has no initial or boundary conditions to simulate them. Furthermore
it is impossible to estimate correctly the parameters that appear in these 
equations 
from the data at hand. To gauge the turbulence strength under these
circumstances one has to proceed indirectly. One well known manifestation
of turbulence strength in the atmosphere is through density fluctuations
(this leads to the well known `twinkling of the stars' phenomena).
Hence turbulence strength is related to the value $C_N^2$ (the refraction
structure constant [8,9]). In the upper troposphere (at
heights of about 10km)
the main contributor to $C_N^2$ is $C_T^2$ -the structure constant for the 
temperature. It follows then that if our conjecture is correct then
the leading Lyaponuv exponent for the temperature time series should 
increase as $C_T^2$ increases.

The plan of the paper is as follows:\,\, In section 2 we introduce the 
six mode truncated model for eqs (\ref{1}) -(\ref{2}) and study the 
behaviors of 
the leading Lyaponuv exponent for the Lorenz model and this system as a 
function of $r$. In section 3 we give a short overview about the atmospheric
structure constants and their computation. In section 4 we consider the time 
series of measurements for the temperature and show that there exist a 
strong correlation between $C_T^2$ and the leading Lyaponuv exponent. 
We end up in section 5 with summary and conclusions.

\setcounter{equation}{0}
\section{Analytical Approach}

In an attempt to relate the leading Lyaponuv exponent  to
the turbulence strength in the flow we consider a truncated expansion 
of eqs.  (\ref{1})-(\ref{2}) with 3 and 6 Fourier components. For the six 
components model we let 
\begin{eqnarray}
\label{21}
\psi &=&
\frac{\sqrt{2}(1+a^2)\kappa}{a}\left\{\left[X_1(t)\sin\left(\frac{\pi
ax}{H}\right) + X_2(t)\sin\left(\frac{2\pi
ax}{H}\right)\right]\right.\\ \nonumber
&&\sin\left(\frac{\pi z}{H}\right)\left.  +
X_3(t)\sin\left(\frac{\pi ax}{H}\right)\sin\left(\frac{2\pi
z}{H}\right)\right\}
\end{eqnarray}
\begin{equation}
\label{2.2}
\theta = \frac{\sqrt{2}\Delta T R_c}{\pi R_a}
\left[Y_1(t)\cos\left(\frac{\pi ax}{H}\right) + Y_2(t)\cos\left(\frac{2\pi
ax}{H}\right)\right]\sin\left(\frac{\pi z}{H}\right) - \frac{\Delta
T R_a}{\pi R_c} Z(t)\sin\left(\frac{2\pi z}{H}\right)
\end{equation}
which leads after some lengthy algebra to the following six equations
for ${X_1,X_2,X_3,Y_1,Y_2,Z}$
\begin{equation}
\label{23}
\dot{X}_1 = \sigma Y_1 - \sigma
X_1+\frac{9}{4}\frac{(a^2-1)\sqrt{2}}{1 + a^2} X_2X_3
\end{equation}
\begin{equation}
\label{24}
\dot{X}_2 = \frac{2\sigma(1+a^2)}{1+4a^2}Y_2 - \frac{\sigma(1+4a^2)}{1+a^2}
X_2 + \frac{9\sqrt{2}}{4(1+4a^2)}\;X_1X_3
\end{equation}
\begin{equation}
\label{25}
\dot{X}_3 = -\frac{\sigma(4+a^2)}{1+a^2}\;X_3 -
\frac{9a^2\sqrt{2}}{4(4+a^2)} X_1X_2
\end{equation}
\begin{equation}
\label{26}
\dot{Y}_1 = \frac{3}{4}\sqrt{2} X_3Y_2 - X_1 Z - Y_1 + rX_1
\end{equation}
\begin{equation}
\label{27}
\dot{Y}_2 = -\frac{3}{4}\sqrt{2} X_3Y_1 - \frac{(1+4a^2)}{1+a^2}Y_2 -
2X_2Z + 2rX_2
\end{equation}
\begin{equation}
\label{28}
\dot{Z} = X_1Y_1 + 2X_2Y_2 - bZ
\end{equation}

For the systems (\ref{5})-(\ref{7}) and (\ref{21})-(\ref{28}) we applied
(the analytical) Wolf 
algorithm [10] to compute the leading Lyaponuv exponent as a function
of $r$. The results of these computations are shown in Fig. 1 for
values of $r$ in the range of [10,115]. We observe that both of these systems 
represent a truncation of the original equations. As a result the flow 
which is represented by eqs.(\ref{1})-(\ref{2}) can not be approximated well by 
the solution of the truncated equations when $r$ is large as additional modes 
become active. As a result the relationship between the solutions of eqs.
(\ref{5})-(\ref{7}) and (\ref{23})-(\ref{28}) and actual convective turbulence 
phenomena is lost as $r$ becomes large. 

From Fig. 1 we see that for both truncated systems under consideration
the leading Lyaponuv exponent remain small or negative for small values
of $r$. It then ``bifurcates'' and goes up sharply at the onset of convective 
state and increases modestly as $r$ is increased further. For the
three mode model as $r$ increases over 70 the modes become saturated 
and the model ceases to approximate the true solution of the convective 
problem. It then follows it own characteristics. 

For the six mode solution, the leading Lyaponuv exponent 
follows the same trajectory as the three mode model up
to $r ~ 70$. However for higher values of $r$ the additional modes in this
model become active and as a result the leading Lyaponuv exponent continue 
to increase up to $r ~ 110$ (where the modes of this model get saturated).

The obvious explanation of this behavior of the leading Lyaponuv exponent
is that the  
the six mode system provides a better approximation to the solution of eqs.
(\ref{1})-(\ref{2}) than the three mode system. As a result the six mode system
captures more of the turbulence phenomena which are related to the solution
of the original system. Accordingly Fig. 1 confirms our conjecture and shows
that the flow in a system with stronger turbulence has also a larger
leading Lyaponuv exponent. 

\setcounter{equation}{0}
\section{Structure Constants}

The structure function of a geophysical variable e.g. the temperature is
defined as

\begin{equation}
D({\bf r})= \langle T^\prime[({\bf r}_1 + {\bf r}),T^\prime ({\bf r}_1)]^{2}\rangle
\end{equation}

where $T^\prime$ are the turbulent fluctuations in the temperature and 
${\bf r}$ is the vector from one point to another.

Kolmogorov showed that for isotropic turbulence in the inertial range this
function depends only on $d$ = $|{\bf r}|$ and scales as

\begin{equation}
D(d) = C_T^{2} d^{2/3} .
\end{equation}

$C_T^2$ which appears as the proportionality constant in this equation
is referred to as the ``temperature structure constant''.

The determination of the atmospheric structure constants [8,9,11,12]
and in particular the temperature structure constant $C_T^2$ is important
in many applications e.g the propagation of electromagnetic signals
[8,9]. Local peaks in the values of these constants, which are indicative 
of strong turbulence and reflect on the structure of the atmospheric flow, 
can have a negative effect on the operation of various optical instruments.

To estimate these structure constants in the upper troposphere or the
stratosphere it is a common practice to send high flying airplanes
that collect data about the basic meteorological variables (such as wind,
temperature and pressure) along its flight path which may extend up
to 200 kms. To estimate the averaged value of the structure constants
along this path one must decompose first the meteorological data into
mean flow, waves and turbulent residuals [13,14]. From the spectrum of the
turbulent residuals one can estimate an averaged value of the structure
constants using Kolmogorov inertial range scaling and Taylor's frozen
turbulence hypothesis [15]. For $C_T^2$ in particular we have
\begin{equation}
            C_T^2(k) = 4 F(k) k^{5/3}
\end{equation}
where $F(k)$ is the temperature spectral density in the inertial range
and $k$ is the wave number. An averaged value for $C_T^2$ (over all wave
numbers in the inertial range) is obtained by averaging these
values over $k$.

In the upper troposphere where humidity is low $C_T^2$ is the main
contributor to the refraction structure constant $C_N^2$ [8,9] which has 
an important impact on the operation of ground telescopes and the 
twinkling of the stars.

\section{Analysis of Atmospheric Time Series}

During 1999 a series of measurements were made over Australia and Japan
by a specially equipped aeroplane [14]. These measurements were taken with 
sampling frequency of 55.1 Hz along a flight path of 200 kms (approximately). 
For each data set the plane flew at almost constant height along a 
straight line at an approximate speed of 103m/sec [14,15].

To use this data to estimate $C_T^2$ we have to split the original
measurements into a sum of mean flow, waves and turbulent residuals.
To accomplish this task we used Karahunen-Loeve(K-L) decomposition
algorithm which has been used by many researchers [15,16].
Actually the last paper applies the K-L decomposition to
the same data considered here and the details of the decomposition
which were given there will not be repeated here.

Based on instrument specifications the data noise should be at a relative
error level of $10^{-3}$. This is confirmed by the eigenvalues obtained
in the K-L decomposition where the last few eigenvalues (which reflect
the noise level in the data) are of order $10^{-3}$ of the leading 
eigenvalue.

For fifteen data sets that were obtained during these flights averaged 
values for $C_T^2$ were obtained using the methodology described above 
[8,9,10].

To compute the leading Lyaponuv exponents for these time series we used
Rosenstein algorithm and its implementation in the TISEAN package [7,17]. 
(However one should note that other algorithms are available for this 
purpose [18]).
To apply this algorithm one has to determine first the `optimal'
delay coordinates and embedding dimension. These were determined using the
mutual information and false neighborhood algorithms [5,19,20] which are 
also implemented in the above mentioned package. This analysis led us to 
choose a four dimensional embedding space with delay coordinates of 600
data points. Fig 2. presents a log-log plot of the results 
for this exponent versus $C_T^2$. We also show on this plot the least
square line for this data. From this plot we see that a change in $C_T^2$
over three orders of magnitudes correlates well with the leading Lyaponuv 
exponents. The fluctuations around the least squares line can be attributed
to wave activity and possible measurements errors. This demonstrates that
the Lyaponuv exponent can be used as a second global measure of turbulence
strength in the data. In cases of discrepancy between $C_T^2$ and 
the leading Lyaponuv exponent one must trace out the reasons for this 
mismatch and correct them.

\section{Summary and Conclusions}

Our main objective in this paper was to link turbulence strength to
the leading Lyaponuv exponent that is related to the flow. This program
was carried out in two contexts. In the first we used a truncated expansion 
of the flow and tested for leading Lyaponuv exponent of the resulting dynamical 
system. In the second part we tested this hypothesis for time series
which represent the flow. In both instances we found a strong supporting 
evidence for our conjecture. Although this does not constitutes a proof
of this conjecture in general we still feel that our results will
useful in many applied contexts especially as a check for the validity of
other global invariants that characterize the flow. In particular the 
determination of the leading Lyaponuv exponent can help verify the value
of $C_T^2$ (or other structure constants) that have important practical 
applications.

\section*{References}

\begin{itemize}

\item[1] Saltzman, B., 1962:\, Finite amplitude free convection as an 
initial value problem, {\it J. Atmos. Sci}, {\bf 19}, 329-341.

\item[2] Lorenz, E.N., 1963:\,  Deterministic nonperiodic flow,
{\it J. Atmos. Sci}, {\bf 20}, 130-141.

\item[3] Marcus, P. S., 1981:\, Effects of truncation in modal representation
of thermal convection, {\it J. Fluid Mech.}, {\bf 103}, 241-255. 

\item[4] Curry, J. H., 1978:\, A generalized Lorenz system, 
{\it Commun. Math. Phys.}, {\bf 60}, 193-204.

\item[5] Abarbanel, H. D. I., Brown, R., Sidorowich, J.J., and Tsimring L.S.,
1993:\, The analysis of observed chaotic data in physical systems,
{\it Rev. Mod. Phys.}, {\bf 65}, 1331.

\item[6] Ott, E., Sauer, T., Yorke, J. A., 1994:\, Coping with Chaos, 
Wiley, NY. 

\item[7] Rosenstein, M. T., Collins, J. J., and De Luca, C. J., 1993:\,
{\it Physica D}, {\bf 65}, 117.

\item[8] Panofsky, H.A., and Dutton, J.A., 1989:\,  Atmospheric
Turbulence,  Wiley, NY.

\item[9] Dewan, E., 1980: \, Optical Turbulence Forecasting,  {\it Air Force
Geophysical Lab.}, AFGL-TR-80-0030, Lab. Report.

\item[10] Fanta, R.L., 1985:\,  Wave Propagation in Random Media,  {\it
Progress in Optics}, {\bf 22}, 342-399.

\item[11] Wolf, A, Swift, J. B., Swinney, H.L., and Vastano, J.A., 1985:\;
Determining Lyapunov exponents from a time series, {\it Physica D},
{\bf 16}, 285-317.

\item[12] Humi, M., 2004:\; Estimation of Atmospheric Structure Constants
from Airplane Data, {\it J. of Atmospheric and Oceanic Technology},
{\bf 21}, 495-500. 

\item[13] Jumper, G.Y. and Beland, R.R., Progress in the understandings and
modeling of atmospheric optical turbulence, {\it AIAA}, 2000-2355,
Proc. 31st AIAA Plasmadynamics and Lasers Conference, 19-22 June 2000,
Denver, CO.

\item[14] Cote, O.R., Hacker J.M., Crawford, T.L., and Dobosy, R.J., 2000:\, 
Clear air turbulence and refractive turbulence in the upper troposphere
and lower stratosphere, Ninth conference on Aviation, Range and Aerospace
Meteorology, Paper No. 16386, Sep 2000.

\item[15] Humi, M., 2003:\, Approximate Simulation of Stratospheric Flow 
from Aircraft Data,
{\it Inter. J. Numerical Methods in Fluids}, {\bf 41}, 209-223.

\item[16] Penland, C., Ghil, M., and Weickmann, K.M., 1991:\, Adaptive filtering
and maximum entropy spectra with applications to changes in atmospheric
angular momentum, {\it J. Geo. Res.}, {\bf 96}, 22659-22671.

\item[17] Hegger, R,  Kantz, H., and Schreiber, T., 1991:\, Practical 
implementation of nonlinear time series methods:\, The TISEAN package, 
Chaos {\bf 9}, 413.

\item[18] Liu, H., Yang, Y., Dai, Z., Yu, Z., 2003:\, The largest Lyapunov
exponent of a chaotic dynamical system in scale space and its
application, {\it Chaos}, {\bf 13}, 839-844.

\item[19] Fraser, A.M. and Swinney, H.L., 1986:\, Independent coordinates for 
strange attractors from mutual information, Phys. Rev. A {\bf 33}, 1134.

\item[20] Kennel, M.B., Brown, R. and Abarbanel, H.D.I., 1992:\, Determining 
embedding dimension for phase-space reconstruction using a geometrical 
construction, Phys. Rev. A, {\bf 45}, 3403.

\item[21] Reiterer P.,Lainscsek C., Schurrer F., Letellier C. and Maquet J.
1998: A nine dimensional Lorenz system to study high dimensional chaos
J. Phys. A {\bf 31} 7121-7139.

\item[22] Frisch U. and Orszag S. -Turbulence: Challenges for theory and 
experiment, Physics today {\bf 43} ,24-29 

\end{itemize}

\newpage

\centerline{\Large{\bf List of Captions}}

Fig. 1: Plot of the leading Lyapunov exponent versus $r=\frac{R_a}{R_c}$
for the Lorenz model(dashed line) and the 6-mode model (solid line)
which was developed in Sec.2.

Fig. 2: Log-log plot of E - the leading Lyapunov exponent- versus $C_T^2$. 
The first order least squares fit for this data yields the relation
$E=10.85*(C_T^2)^(0.2760)$ .

\newpage

\begin{figure}[htb]
\centering
\centerline{\psfig{file=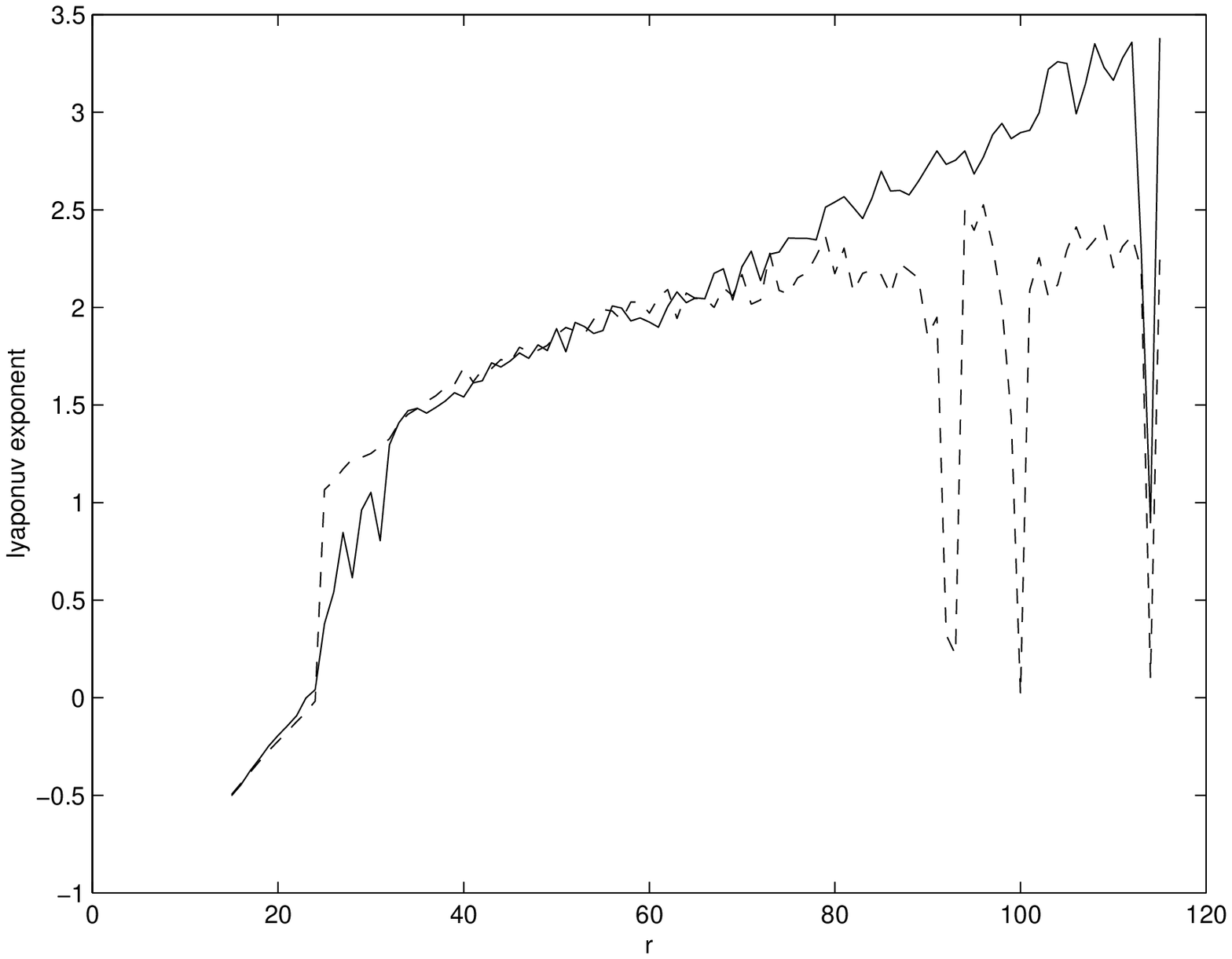,height=4in,width=5.5in}}
\caption{}
\end{figure}

\newpage

\begin{figure}[htb]
\centering
\centerline{\psfig{file=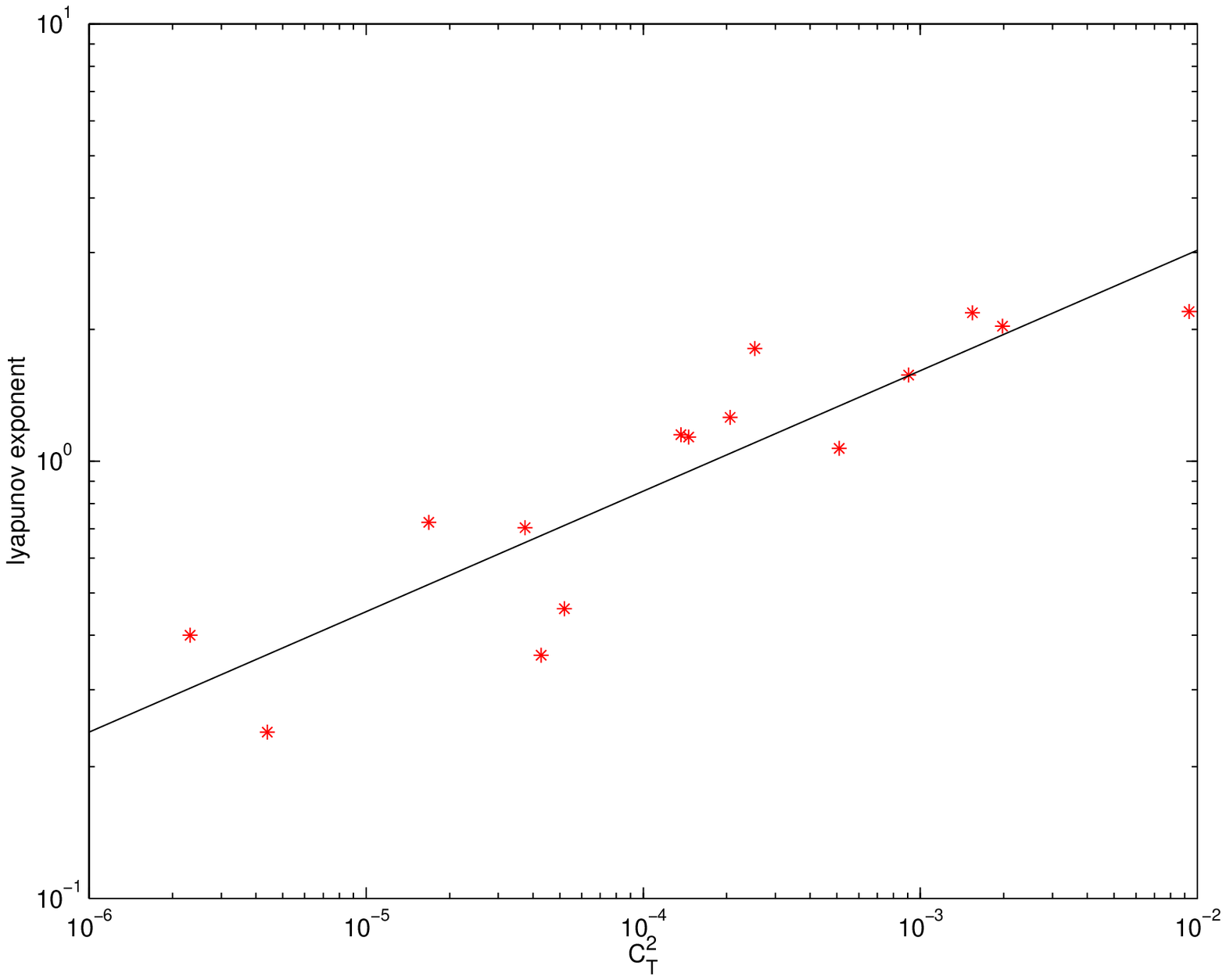,height=4in,width=5.5in}}
\caption{}
\end{figure}

\end{document}